# AN IMPROVED DEEP NEURAL NETWORK FOR MODELING SPEAKER CHARACTERISTICS AT DIFFERENT TEMPORAL SCALES

*Bin Gu, Wu Guo*

National Engineering Laboratory for Speech and Language Information Processing,
University of Science and Technology of China, Hefei, China

**ABSTRACT**

This paper presents an improved deep embedding learning method based on convolutional neural network (CNN) for text-independent speaker verification. Two improvements are proposed for x-vector embedding learning: (1) Multi-scale convolution (MSCNN) is adopted in frame-level layers to capture complementary speaker information in different receptive fields. (2) A Baum-Welch statistics attention (BWSA) mechanism is applied in pooling-layer, which can integrate more useful long-term speaker characteristics in the temporal pooling layer. Experiments are carried out on the NIST SRE16 evaluation set. The results demonstrate the effectiveness of MSCNN and show the proposed BWSA can further improve the performance of the DNN embedding system.

## 1. INTRODUCTION

Speaker verification (SV) is a process of verifying a person's claimed identity from some speech signal. Over the years, the i-vector [1] algorithm has achieved significant success in modeling speaker identity and channel variability, in which an utterance is represented in the form of a fixed- and low-dimensional feature vector. Combined with an effective backend classifier, the probabilistic linear discriminant analysis (PLDA) [2], the i-vector/PLDA framework has been the dominant approach for the last decade.

With the great success of deep learning over a wide range of machine learning tasks, more efforts have been focused on using deep neural network (DNN) to extract more discriminative speaker representations [3,4,5,6]. These deep speaker embedding systems can achieve comparable or even better performance compared with the i-vector based methods, particularly under conditions of short-duration utterances.

Since the time delay neural network (TDNN) based x-vector system was proposed [32], a series of similar improved networks have been investigated [18,19,25], and a variety of other neural network architectures, such as DenseNet, ResNet and InceptionNet, have also been used to extract more speaker-discriminative embeddings [21,23,24]. In order to obtain long-term speaker representation with more discriminative power, attention mechanism [17] is widely used recently. In [27], attentive statistics pooling was proposed to replace the conventional statistics pooling. In [10], multi-head self-attention mechanism was applied. In [28], the authors explored different topologies and their variants of attention layer, and compared different pooling methods on attention weights.

The above-mentioned x-vector and its variants have achieved great success in SV field, but there still exists room for improvement. From input layer towards output layer, a typical x-vector system includes stacked frame-level layers, pooling layer and utterance-level layers. Most systems have only one set of filters with single receptive field in a frame-level layer. Training a model where multiple filters with varying window sizes are applied in each layer is an effective and widely used method in the natural language processing (NLP) task [11, 12, 13]. We applied multi-scale convolution (MSCNN) in frame-level layers, and found different receptive fields can provide complementary ability for catching speaker characteristics. Through a linear combination of features from different receptive fields, we obtained more robust frame-level representations. Though attention mechanism equips x-vector system with the ability to focus on the important frames in speaker modelling, the conventional self-attention method can't fully utilize the inner relationships between the frames and the utterance. To address this problem, i-vectors are applied in the pooling layer with attention mechanism in [10]. Extracting i-vectors from acoustic features can be looked on as a dimensionality reduction process, and this will lose some strength in catching the speaker discriminative ability. We resorted to integrate the Baum-Welch statistics into attention mechanism in this work, and found Baum-Welch statistics based attention (BWSA) more effective than the i-vector based attention. The experimental results on the NIST SRE16 evaluation set demonstrate the power of the proposed methods.

The remainder of this paper is organized as follows. Section 2 describes our x-vector baseline system and the attentive statistics pooling. Section 3 introduces the proposed methods. The experimental setup, results and analysis are presented in Section 4. Finally, the conclusion is given in Section 5.

## 2. BASELINE NETWORK ARCHITECTURE
### 2.1. X-vector baseline system
The network architecture of our baseline x-vector system is the same as that described in [8].

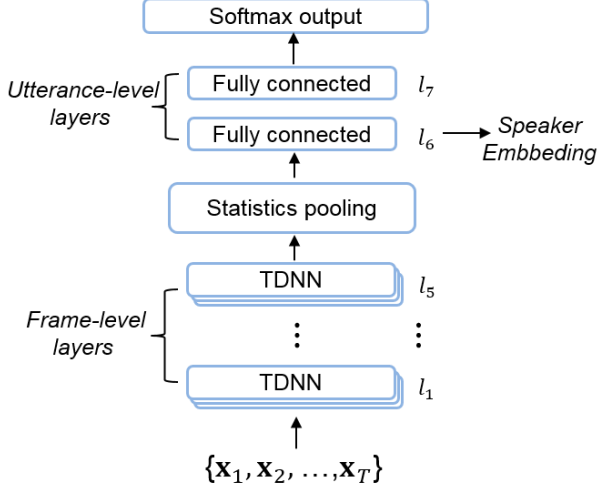

**Figure1:** *Network architecture of the x-vector baseline*

As depicted in Figure 1, five TDNN layers are stacked as the frame-level feature extractor, and the final output vectors of the whole variable-length utterance are aggregated into a fixed utterance-level vector through the statistics pooling. The mean and standard deviation are calculated and then concatenated together as the output of the pooling layer. Two additional fully connected layers $l_6$ and $l_7$ are followed to obtain a low-dimensional speaker representation after the pooling layer and finally passed into a softmax output layer.

### 2.2. Attentive statistics pooling
The attention mechanism offers a way to select the more relevant frame-level representations than others in the input utterance that better expresses speaker characteristics.

For an utterance with $T$ frames, the attention weight $\alpha_t$ for the $t$-th output representation $\mathbf{h}_t \in \mathbb{R}^D$ of the last frame-level layer can be calculated as

$$e_t = f(\mathbf{h}_t), t = 1, ..., T \quad (1)$$

$$\alpha_t = \frac{g(e_t)}{\sum_{k=1}^{T} g(e_k)} \quad (2)$$

where $f(\cdot)$ is a non-linear projection that converts a vector to a scalar score and $g(\cdot)$ is a function that ensures the condition $\alpha_t \geq 0$ holds. Function $g(\cdot)$ is usually an exponential function or a logistic function.

Then the mean $\boldsymbol{\mu} \in \mathbb{R}^D$ and standard deviation $\boldsymbol{\sigma} \in \mathbb{R}^D$ of weighted representations $\{\alpha_1 \mathbf{h}_1, \alpha_2 \mathbf{h}_2, ..., \alpha_T \mathbf{h}_T\}$ are concatenated as the output $\mathbf{c} \in \mathbb{R}^{2D}$ for the attentive statistics pooling layer:

$$\mathbf{c} = [\boldsymbol{\mu}, \boldsymbol{\sigma}] \quad (3)$$

The attention mechanism varies with different key vectors and functions $f(\cdot)$ in Eq. (1). The self-attention and i-vector based attention stand for two kinds of algorithm in SV field [9, 10].

- In single-head self-attention [9], Eq. (1) can be written as:

$$e_t = f_{SA}(\mathbf{h}_t) = f_{SA}(\mathbf{h}_t, [\mathbf{w}_1, ..., \mathbf{w}_n, ..., \mathbf{w}_N]^T)$$
$$= \mathbf{v}^T \tanh(\mathbf{W}\mathbf{h}_t + \mathbf{b}) \quad (4)$$

where $\mathbf{v} \in \mathbb{R}^N$, $\mathbf{W} \in \mathbb{R}^{N \times D}$ and $\mathbf{b} \in \mathbb{R}^N$ are trainable parameters. These parameters keep fixed for the whole corpus, and the attention weights are all determined by $\mathbf{h}_t$.

- In i-vector based attention [10], Eq. (1) are modified as:

$$e_t = f_{IA}(\mathbf{h}_t) = f_{IA}(\mathbf{h}_t, \mathbf{r}) = \frac{\mathbf{h}_t \odot \mathbf{r}}{\|\mathbf{h}_t\|_2 \|\mathbf{r}\|_2} \quad (5)$$

where $\odot$ is the elementwise multiplication, and $\mathbf{r} \in \mathbb{R}^D$ is the transformed i-vector through a non-linearly affine layer. Since each utterance owns an utterance-dependent i-vector, the attention weights are determined by both $\mathbf{h}_t$ and the utterance-level information from the i-vector.

## 3. PROPOSED METHOD
In this section, we improve the frame-level and pooling-layers of x-vector system to extract more discriminative speaker embedding. Firstly, convolution with multi-scale filters is used in frame-level layers to provide complimentary temporal information. Secondly, Baum-Welch statistics is adopted in attentive pooling layer to incorporate extra utterance-level information.

### 3.1. Convolution with multi-scale filters
Different from most current SV systems that only implement a single set of filters in a convolution layer, we develop a convolutional network module that aggregates multi-scale context information in this work. To accelerate the training efficiency, we employ a few sets of 1-D convolution filters in several frame-level layers. Meanwhile, depthwise separable convolution [14] is used to replace the regular convolution.

Suppose the output of the $l^{th}$ frame-level layer by $\mathbf{H}_l \in \mathbb{R}^{T \times C}$, where $C$ is the number of channels for convolution. For the $l+1^{th}$ layer, $K$ sets of convolution filters $\{\mathbf{W}_{l+1}^1, ..., \mathbf{W}_{l+1}^K\}$ with various dilation factors are applied separately.

$$\mathbf{s}_{l+1}^c = \text{relu}(\mathbf{W}_{l+1}^k * \mathbf{H}_l + \mathbf{b}), c \in [\lambda(k-1), \lambda k] \quad (6)$$

$$\mathbf{H}_{l+1} = [\mathbf{s}_{l+1}^1, ..., \mathbf{s}_{l+1}^C] \quad (7)$$

where $*$ indicates the convolution operation and $\lambda$ equals $C/K$. 1-dimentional vectors $\mathbf{s}_l^1, ..., \mathbf{s}_l^C \in \mathbb{R}^T$ can be viewed as $C$ temporal sequences from different channels. In a word, the output $\mathbf{H}_{l+1}$ consists of $K$ parts, and they are computed from the output $\mathbf{H}_l$ with $K$ sets of filters working in parallel.

## 3.2. BWSA based statistics pooling

The Gaussian mixture model-universal background model (GMM-UBM) can map the variable length speech file into a fixed-dimensional matrix [16, 17].

$$\mathbf{f}_m = \sum_t \gamma_t(m)\mathbf{x}_t / T, m = 1,...,M \quad (8)$$

$$\mathbf{F} = [\mathbf{f}_1,...,\mathbf{f}_m,...,\mathbf{f}_M] \quad (9)$$

where $M$ is the total Gaussian components, $\gamma_t(m)$ denotes the posterior probability of acoustic feature $\mathbf{x}_t$ against the $m^{th}$ Gaussian component, and $\mathbf{F}$ is a matrix of the normalized first-order statistics. Different from the i-vector, $\mathbf{F}$ has no information loss.

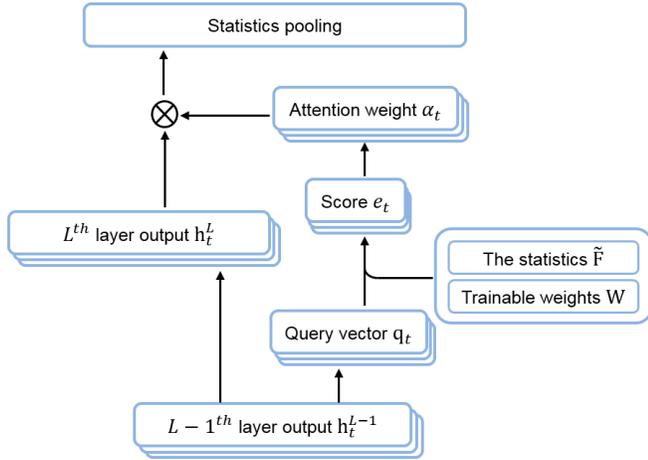

**Figure2:** *Proposed attentive statistic pooling*

The proposed BWSA based statistics pooling is depicted in Fig. 2. Suppose that there are $L$ frame-level layers below the pooling layer, a low-dimension query vector $\mathbf{q}_t$ is generated from the output $\mathbf{h}_t^{L-1}$:

$$\mathbf{q}_t = d(\mathbf{h}_t^{L-1}) \quad (10)$$

where $d(\cdot)$ is a non-linear function.

Both the Baum-Welch statistics and a trainable weight matrix are concatenated as the key vector matrix.

$$\tilde{\mathbf{f}}_m = \mathbf{V}_2 \tanh(\mathbf{V}_1 \mathbf{f}_m + \mathbf{b}) \quad (11)$$

$$\mathbf{K} = [\tilde{\mathbf{f}}_1,...,\tilde{\mathbf{f}}_m,...,\tilde{\mathbf{f}}_M, \mathbf{w}_1,...,\mathbf{w}_n,...,\mathbf{w}_N]^T$$
$$= [\tilde{\mathbf{F}}, \mathbf{W}]^T \quad (12)$$

where $\mathbf{V}_1$, $\mathbf{V}_2$ and $\mathbf{W}$ are trainable parameter matrix. Note $\tilde{\mathbf{f}}_m$ should have the same dimension with $\mathbf{w}_n$.

The attention score can be calculated as follows.

$$e_t = f_{BA}(\mathbf{h}_t^{L-1}) = \mathbf{v}^T \tanh(\mathbf{K}\mathbf{q}_t + \mathbf{b}) \quad (13)$$

Then the convention steps in Eq (2) and (3) are applied for the following statistics pooling.

## 4. EXPERIMENTS

### 4.1. Data set and evaluation metric

The experiments are conducted on the NIST SRE16 evaluation set, which includes Tagalog and Cantonese telephone speech. The duration of enrollment data is around 60 seconds, and that of test speech is ranging approximately from 10 seconds to 60 seconds. In addition, an unlabeled set of 2272 recordings in the development set is available for training models such as GMM-UBM. The training data primarily consists of the telephone speech from past NIST SRE (2004-2010), Switchoard and Mix6 datasets. Data augmentation techniques described in [8] including adding additive noise and reverberation data are applied to improve the robustness of the system. Totally, there are 207759 recordings from 6986 speakers. The performance is reported in terms of equal error rate (EER) as well as the minimal detection cost function (minDCF).

### 4.2. Features

60-dimensional MFCC features containing delta and delta-delta coefficients extracted from the speech of a 25ms window with 10ms frame shift are used for the i-vector system, while the input acoustic features for deep speaker embedding systems are 23-dimentional MFCC features. These features are mean-normalized over a 3 second sliding window, and energy based VAD is employed to filter out non-speech frames.

### 4.3. Front-end embedding extractors

Six front-end embedding extractors, one i-vector and five x-vector systems, are investigated in this work.

*4.3.1. I-vector baseline*

The features extracted from the unlabeled utterances of the development set are used to train a 512-component diagonal-covariance GMM-UBM. After training a total variability space loading matrix, 400-dimensional i-vectors can be obtained.

*4.3.2. X-vector baseline*

This is the baseline x-vector system [8] implemented with the TensorFlow toolkit [33]. Each of the bottom four frame-level layers has 512 nodes, while there are 1500 hidden nodes in the fifth layer. The dilation factors of these five layers are 1,2,3,1 and 1 respectively. Both of the two fully connected layer after statistic pooling layer have 512 nodes. Batch normalization is used on all layers expect the attentive statistics pooling layer. In order to prevent from overfitting, the dropouts as well as L2-weight decay are employed in the system. The network is trained on 10-second chunks. We utilize Adam optimizer to train the network. The initial learning rate is 0.00015, and it is decayed based on the validation set.

### 4.3.2. Four improved deep speaker embedding systems

Four improved speaker embedding systems are experimented. Besides the proposed methods, two contrastive systems are also compared. These four systems have almost the same structure as the baseline system except for the modification parts as stated as follows.

**SA**: The self-attention described in Section 2.2 is applied in this contrastive system. The number of hidden nodes in the attentive statistic pooling layer is set to 512.

**IA**: This system is also a contrastive system which uses the i-vector based attentive statistics pooling [10].

**BA**: The proposed BWSA based statistics pooling is employed in this system. The dimensionality of trainable weight vectors in Eq. (12) is 512.

**BA+MS-3L**: Besides the BWSA, MSCNN are adopted in the bottom three frame-level layers. Two sets of convolution kernels are used. The first set has the same kernel size and dilation rate with those in baseline system, while the other only doubles the dilation rate.

### 4.4. Backend classifier

After extracting i-vectors or x-vectors, the evaluation set are centered using the unlabeled development set. The dimensions of the vectors are reduced to 200 through LDA algorithm. Data whitening and length normalization are applied before PLDA. After these pre-processing steps, the PLDA model with unsupervised adaptation is trained and used as backend classifier for speaker verification. The backend classifier is implemented with the Kaldi toolkit [31].

### 4.5. Results and analysis

Table 1 *Comparison results of different systems on SRE16*

| system | Pooled | | Tagalog | | Catonese | |
|---|---|---|---|---|---|---|
| | EER | $DCF^{min}$ | EER | $DCF^{min}$ | EER | $DCF^{min}$ |
| i-vector | 14.08 | 0.739 | 17.31 | 0.864 | 8.20 | 0.597 |
| x-vector | 7.99 | 0.587 | 11.58 | 0.741 | 4.26 | 0.430 |
| SA | 7.61 | 0.575 | 11.04 | 0.729 | 4.23 | 0.423 |
| IA | 7.81 | 0.586 | 11.15 | 0.736 | 4.54 | 0.437 |
| BA | 7.29 | 0.569 | 10.74 | 0.733 | 3.88 | 0.402 |
| BA+MS-3L | **7.04** | **0.561** | **10.34** | **0.725** | **3.77** | **0.398** |

Table 1 presents the performance of different systems on SRE16 test set. It can be observed that all the systems with attention mechanisms outperform two baseline systems, which demonstrates the effectiveness of attention mechanisms in SV task. Among the three different attention mechanic (SA, IA and BA), the proposed Baum-Welch statistics attention mechanism achieves best performance and IA only achieves marginal improvement over the baseline. The reason may be that Baum-Welch statistics has kept all the original utterance-level information without any loss. Further improvement can be observed when the MSCNN is used, and the BA+MS-3L can get 12% EER reduction compared with the x-vector baseline.

Table 2 *Results with different system configurations. "L" indicates the number of layers applying MSCNN. "K" is the parameter described in Section 3.1. "N" denotes the MSCNN layer size.*

| system | L | K | N | Pooled | |
|---|---|---|---|---|---|
| | | | | EER | $DCF^{min}$ |
| x-vector | - | - | 512 | 7.99 | 0.587 |
| x-vector* | - | - | 756 | 8.11 | 0.596 |
| MS-1L | 1 | 2 | 512 | 7.88 | 0.589 |
| MS-2L | 2 | 2 | 512 | 7.65 | 0.575 |
| MS-3L | 3 | 2 | 512 | 7.60 | 0.572 |
| MS-3L* | 3 | 3 | 756 | **7.51** | **0.571** |

In the following experiments, we investigate the effect of the MSCNN with different configurations, and the x-vector system without attention strategy are adopted for simplicity. MS-1L, MS-2L and MS-3L stand for the systems with bottom one, two and three frame-level layers using the MSCNN respectively. As listed in Table 2, the more layers employ the MSCNN, the better performance achieves. These multi-scale filters with various receptive-field size can provide complementary speaker characteristics at different temporal scales.

We also explore the influence of using more sets of filters in MSCNN layers. Because there is an obvious performance degradation if we just use bigger $K$ with the hidden layer size fixed, we build a system (MS-3L*) where the first three frame-level layers have 756 hidden nodes respectively and $K$ is set to 3. For fairly comparison, we also build another x-vector* system having the same hidden layer size with MS-3L*. From the results in Table 2, the system equipped with more MSCNNs (the last row) can achieve the best EER, and this observation proves more convolution filters can provide extra complementary information.

## 5. CONCLUTION

In this paper, we investigate the complementarity of speaker information at different temporal scales. More specifically, multi-scale filters with different receptive filed are introduced in convolutional layers. In this way, the information of different granularity at frame level can be detected. In order to capture the utterance-level speaker information better, a Baum-Welch statistics attention mechanism is proposed, where both local fixed key vectors (the statistics) and the global fixed key vectors (the weight vectors initialized randomly) are utilized. In our future study, we will continue to investigate more useful methods to highlight both the frame- and utterance- level speaker characteristic for x-vector systems.

## 6. ACKNOWLEDGEMENTS

This work was partially funded by the National Natural Science Foundation of China (Grant No. U1836219) and the National Key Research and Development Program of China (Grant No. 2016YFB100 1303).